\documentclass[aps,prl,twocolumn,groupedaddress,amsmath]{revtex4}

\usepackage{graphicx}

\begin{document}

\preprint{PRL}


\title{High-Frequency Trading Synchronizes Prices in Financial Markets}



\author{Austin Gerig\\austin.gerig@sbs.ox.ac.uk\\CABDyN Centre, Sa\"{i}d Business School, University of Oxford, United Kingdom}



\begin{abstract}

High-speed computerized trading, often called ``high-frequency trading'' (HFT), has increased dramatically in financial markets over the last decade. In the US and Europe, it now accounts for nearly one-half of all trades.  Although evidence suggests that HFT contributes to the efficiency of markets, there are concerns it also adds to market instability, especially during times of stress.  Currently, it is unclear how or why HFT produces these outcomes.  In this paper, I use data from NASDAQ to show that HFT synchronizes prices in financial markets, making the values of related securities change contemporaneously.  With a model, I demonstrate how price synchronization leads to increased efficiency: prices are more accurate and transaction costs are reduced.  During times of stress, however, localized errors quickly propagate through the financial system if safeguards are not in place.  In addition, there is potential for HFT to enforce incorrect relationships between securities, making prices more (or less) correlated than economic fundamentals warrant.  This research highlights an important role that HFT plays in markets and helps answer several puzzling questions that previously seemed difficult to explain: why HFT is so prevalent, why HFT concentrates in certain securities and largely ignores others, and finally, how HFT can lower transaction costs yet still make profits.

\end{abstract}

\pacs{}


\maketitle


\section{Introduction}

Over the past 10 years, high-frequency trading (hereafter HFT) has gone from a small, niche strategy in financial markets to the dominant form of trading.  It currently accounts for approximately 55\% of trading volume in US equity markets, 40\% in European equity markets, and is quickly growing in Asian, fixed income, commodity, foreign exchange, and nearly every other market\footnote{Several research firms provide estimates of HFT activity for subscribers; examples are the TABB Group, the Aite Group, and Celent.  Publicly, this information is available in articles such as ``The fast and the furious'', Feb. 25, 2012,  \emph{The Economist} and ``Superfast traders feel the heat as bourses act'', Mar. 6, 2012, \emph{Financial Times}.}.  Although a precise definition of HFT does not exist, it is generally classified as autonomous computerized trading that seeks quick profits using high-speed connections to financial exchanges.  

Policy makers across the globe are spending considerable effort deciding if and how to regulate HFT\footnote{See for example the SEC document ``Concept Release on Equity Market Structure'' available at www.sec.gov, the ESMA document ``Guidelines on systems and controls in a highly automated trading environment for trading platforms, investment firms and competent authorities'' available at www.esma.europa.eu, the European Commission document ``Consultation on financial sector taxation'' available at ec.europa.eu, and the BIS Foresight project ``The Future of Computer Trading in Financial Markets'' available at www.bis.gov.uk.}  On the one hand, HFT appears to make markets more efficient.  Algorithmic trading in general, and HFT specifically, increases the accuracy of prices and lowers transaction costs\cite{Hendershott11a, Litzenberger12, Hendershott11c, Menkveld12}.  On the other hand, HFT appears to make the financial system as a whole more fragile.  The rapid fall and subsequent rise in prices that occurred in US markets on May 6, 2010 (known as the ``Flash Crash''), was, in part, due to HFT\cite{Kirilenko11}.  Because HFT firms do not openly disclose their trading activities, it has so far been unclear how and why HFT produces these outcomes; a circumstance that has greatly increased the controversy surrounding its existence.  

In this paper, using a special dataset supplied by NASDAQ, I present evidence that HFT synchronizes security prices in financial markets.  By `synchronize', I mean the following -- to the extent that two securities are related to one another, HFT activity ensures that a price change in the first security coincides nearly instantaneously with a similar price change in the second security.  Synchronization is a gargantuan task\footnote{There are over one thousand transactions per second in US equities alone during the trading day (see ``U.S. Consolidated tape Data'' available at www.utpplan.com).} tailor-made for HFT: it is profitable for the firms that do it and can only be done with high-speed computerized trade.

To understand the effects of price synchronization, I modify a standard model of price formation\cite{Glosten85} so that it includes multiple related securities.  I find that when prices are synchronized, transaction costs are reduced, prices are more accurate, and that informed investors -- those who always submit a buy (sell) order when the price will be higher (lower) -- make less profits.

The intuition behind these results is straightforward.  As an example, suppose that an event occurs which increases the likelihood that country X will default on its sovereign debt.  This information is processed by specialized firms who quickly buy securities that track the probability of X's default.  The prices of these securities increase, and if markets are synchronized, then the prices of all other securities adjust as well.  As a result, an investor who purchases or sells any security in the market receives a more accurate price.  Transaction costs are reduced because liquidity providers are more confident in market prices and require less of a price concession to transact with an order.  In finance, this is known as a reduction in adverse selection costs\cite{Biais05,Baruch09,Gerig10}.

\begin{figure*}
\begin{center}
\includegraphics[width=6.8in]{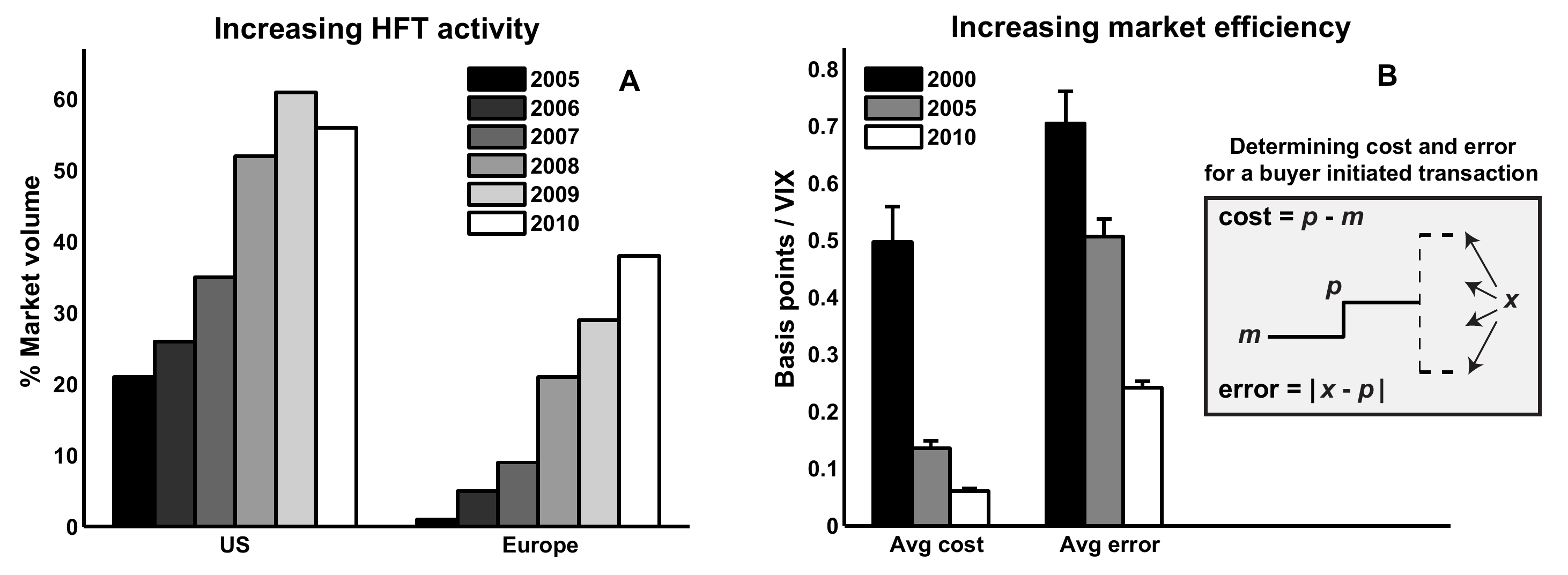}
\end{center}
\caption{Increase of HFT activity and market efficiency over the last decade.  (A) The percentage of HFT volume in US equities (measured in shares) and European equities (measured in value) for 2005 to 2010.  Values are estimates from the TABB Group. (B) Average cost and average pricing error of transactions for 35 US stocks during the last full week of February in 2000, 2005, and 2010.  The diagram shows how the cost and error of a buyer initiated transaction are calculated (for a seller initiated transaction, the sign of the cost is reversed).   To normalize across stocks and time, costs and errors are measured in basis points (1bps=.01\%) and divided by the current market volatility as measured by the VIX index.  Error bars report the standard error of the mean across the 35 stocks.}
\end{figure*}

If transaction costs are lower, then average investors benefit from synchronization.  So, who loses?  When prices are synchronized, information diffuses rapidly from security to security and informed investors are made somewhat redundant.  In the model, they make less profit as a result.

Although price synchronization is normally beneficial in markets, it can also have harmful effects.  When prices are tightly connected to one another, localized errors quickly propagate through the financial system.  In addition, there is potential for incorrect relationships between securities to be enforced, making prices more (or less) correlated than economic fundamentals warrant.  Finally, during times of market stress, HFT firms are impelled to leave the market if their systems observe events outside the parameters they are programmed to handle -- a circumstance that causes liquidity to disappear at the precise time it is needed the most.

\begin{figure*}
\begin{center}
\includegraphics[width=6.8in]{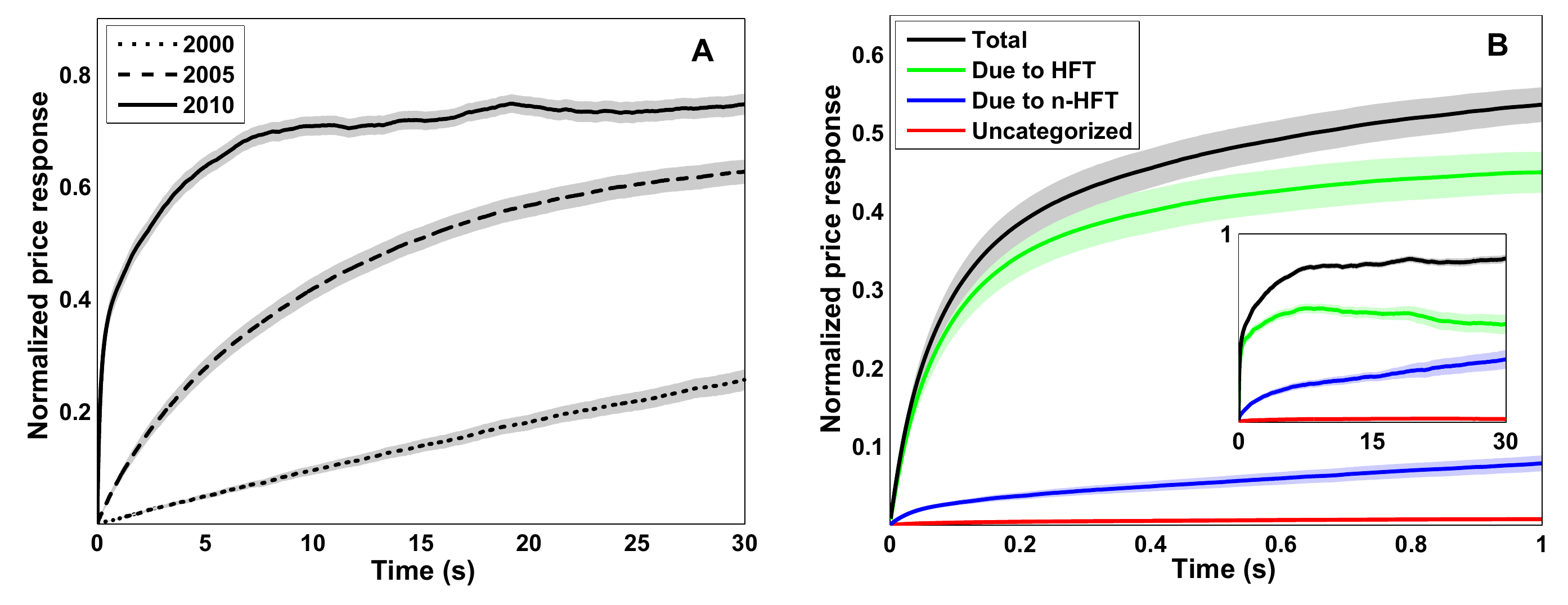}
\end{center}
\caption{Normalized price response of stock $i$ to a price movement in stock $j\neq i$.  (A) The mean price response for 35 US stocks during the last full week of February in 2000, 2005, and 2010.  The standard error of the mean across the 35 stocks is shown in gray.  (B) For 2010, the price response of the full 40 US stocks (black) is decomposed into the amount due to HFT activity (green), non-HFT activity (blue), and an amount that could not be categorized (red).  Again, standard errors are shown in shaded color.}  
\end{figure*}

\section{Evidence}

In Fig.~1, I show the rapid increase in HFT volume and the corresponding increase in market efficiency over the last decade.  In Fig.~1(A), HFT estimates are from the TABB Group.  In Fig.~1(B), the cost of a buyer initiated transaction is measured as the transaction price, $p$, minus the current prevailing midpoint price for the security, $m$, (for a seller initiated transaction, the cost is $m-p$).  The transaction error is the absolute difference between the transaction price, $p$, and the midpoint price 1 minute later, $x$, (see the diagram in the figure).  Data is from Thomson Reuters and includes 35 stocks during the last full week of February in 2000, 2005, and 2010.  The 35 stocks are a subset of the 40 large-cap stocks available in the NASDAQ HFT data that is used below (5 of the stocks from the NASDAQ data are not available in the Reuters data for all time periods). 

In Fig.~2(A), I show how prices have become more synchronized over the last decade.  Using the same data as in Fig.~1(B), I measure the average normalized price response of security $i$ to a price movement in security $j\neq i$ (see the supplemental material for details\cite{Materials}).  In 2000, it took several minutes for a price movement in stock $j$ to be fully incorporated into the price of stock $i$.  In 2005, this occurred in about 1 minute, and in 2010 it took less than 10 seconds.  
Fig.~2(B) shows that it is HFT activity that keeps prices synchronized.  Using data from NASDAQ that flags HFT activity in 120 stocks during the last week of February, 2010 (see the supplemental material for a full description of the data\cite{Materials}), I take the 40 largest stocks and calculate the average normalized price response of stock $i$ to a price movement in stock $j$ (black curve).  I separate this response into the amount due to HFT activity (green curve), non-HFT activity (blue curve), and an amount that could not be categorized either way (red curve).  As seen in the figure, an overwhelming majority of the initial price response is due to HFT activity.

\section{Model}

To study the effects of price synchronization in detail, I modify a standard model of price formation\cite{Glosten85} so that it includes multiple related securities.  In total $n$ securities, $i=1,2,\dots,n,$ exist and are traded asynchronously over a single period.  During the trading period, one unit-sized order to buy or sell is submitted for each security, $O_i\in\{B_i, S_i\}$.  Submitted orders are immediately transacted by liquidity providers at the fair price, i.e., at the expected future price of the security, $p_i$.  The original price of security $i$ is $m_{i}$, and the final price at the end of trading, $x_{i}$, increases or decreases with equal probability, $x_i\in\{x_i^+,x_i^-\}$, where $x_{i}^+ = m_{i}+\delta_i$ and $x_{i}^- = m_{i}-\delta_i$, with $\delta_i>0$.  So in summary, for each security $i$, one order is submitted, transacts at price $p_i$, and the final price is either $\delta_i$ higher or $\delta_i$ lower than the original price.

In real markets, informed individuals correlate their orders with future price changes, i.e., buying tends to correspond with increases in prices and selling with decreases in prices.  To include this effect, I assume that a buy order is more likely when the final price of the security is higher, and that a sell order is more likely when it is lower.  For security $i$, $\mathcal{P}(B_i|x^+_{i})=\phi_i>0.5$ and $\mathcal{P}(B_i|x^-_{i})=(1-\phi_i)<0.5$.  Finally, in real markets, securities are related to one another so that their price changes are correlated.  To include this effect, I assume that the price change of security $i$ and $j$ are correlated with correlation coefficient, $\rho_{i,j}\neq0$.   To complete the model, I assume that the orders for securities are independent of one another, except through the indirect dependence caused by the correlations already assumed, $\mathcal{P}(O_1,O_2,\dots|x_{1},x_{2},\dots)=\mathcal{P}(O_1|x_{1})\mathcal{P}(O_2|x_{2})\dots$.

\begin{figure*}
\begin{center}
\includegraphics[width=6in]{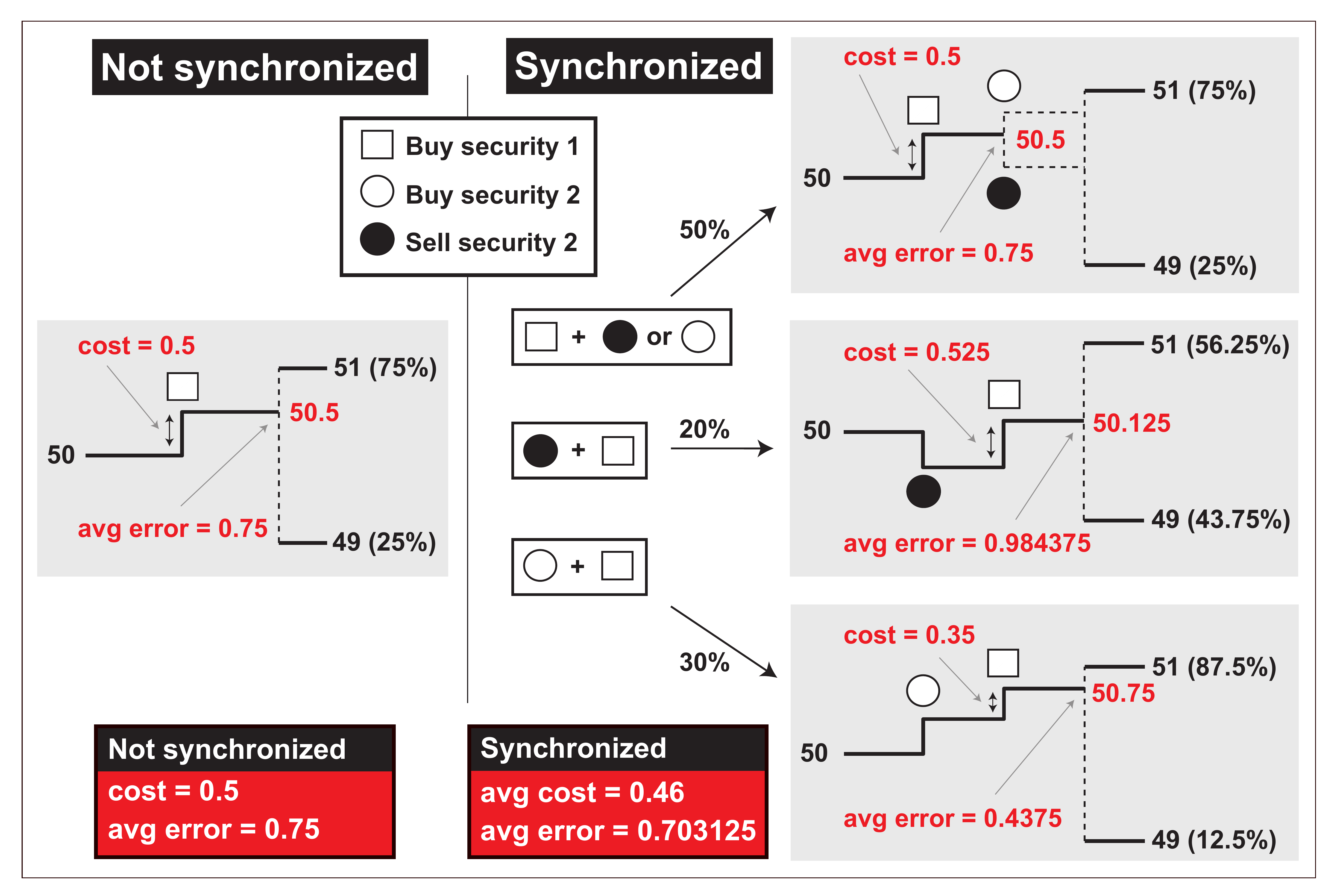}
\end{center}
\caption{Diagram of the model showing that transaction costs are reduced and prices are more accurate with synchronized prices.  Only a buy order for security 1 is analyzed.  On the left, the price of security 1 and 2 are not synchronized.  The buy order for security 1 transacts at 50.5, giving a transaction cost of 0.5 and an average transaction price error of 0.75.  On the right, the prices are synchronized.  If the buy order for security 1 arrives before the order for security 2 (top), the analysis is the same as if prices where not synchronized.  If a buy or sell order for security 2 arrives first (middle or bottom), this affects the price of security 1 as shown.  Transaction costs and errors are calculated for each case and averaged.  Final results are shown at the bottom in reverse red highlight.}
\end{figure*}

\begin{figure*}
\begin{center}
\includegraphics[width=6.8in]{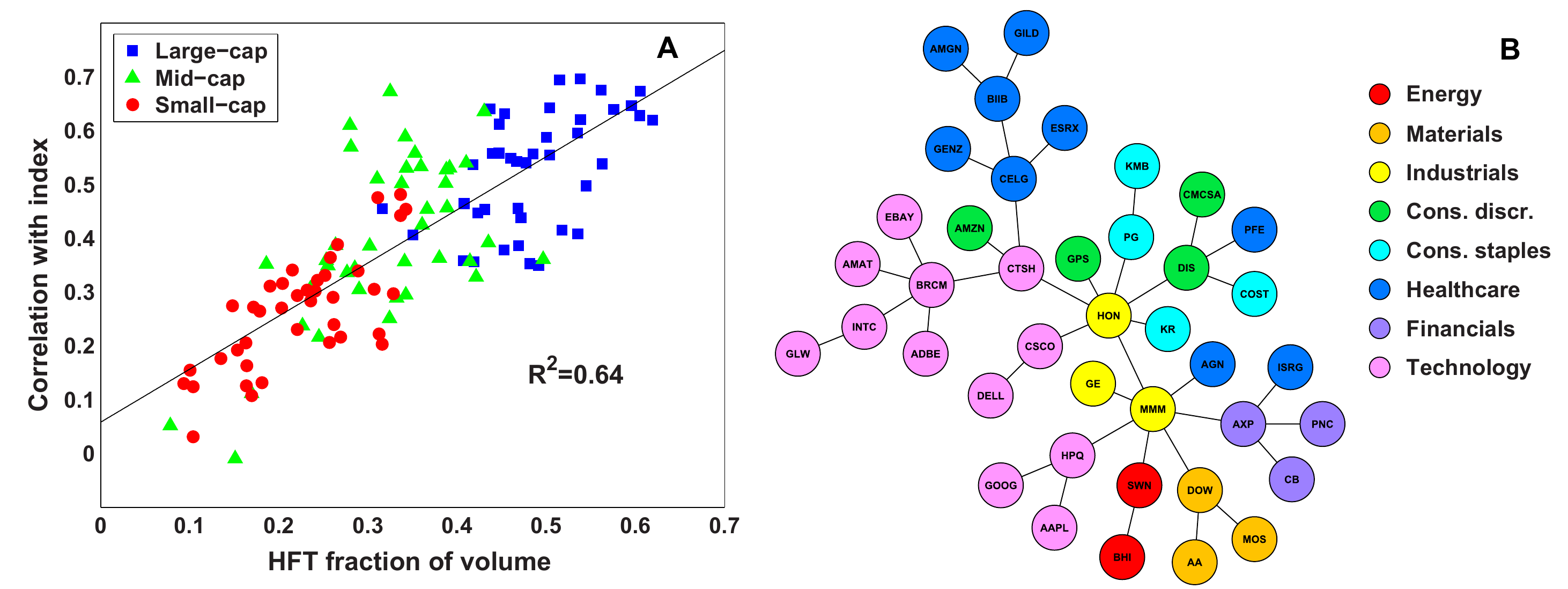}
\end{center}
\caption{Stocks with higher HFT activity have stronger correlations with other stocks.  These correlations correspond with economic structure.  (A) Plot of stock correlation vs. the fraction of volume due to HFT for that stock.  Correlations are between the 30 second returns of the stock and the equal-weighted average 30 second returns of all 120 stocks.  Volume is measured in shares.  (B) Minimum spanning tree derived from the 30 second correlation matrix for the 40 large-cap stocks.  The ticker for each stock is shown on the corresponding node, and nodes are color-coded according to GICS sector.}
\end{figure*}

\begin{figure*}
\begin{center}
\includegraphics[width=6.8in]{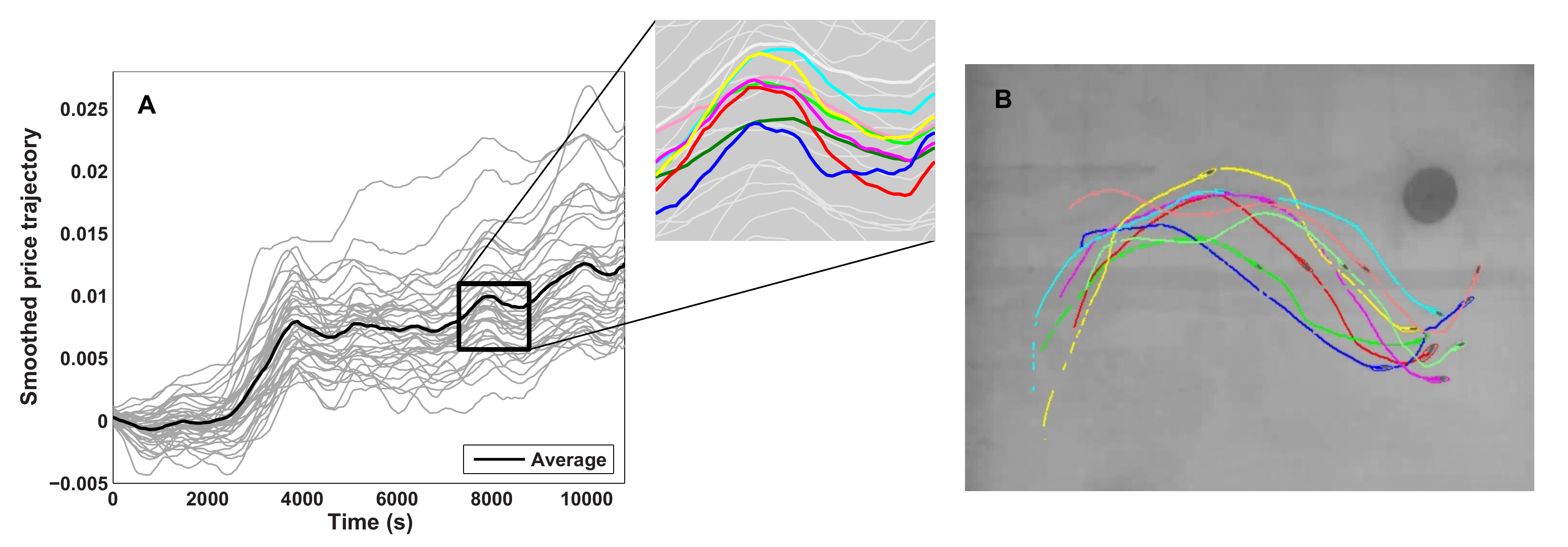}
\end{center}
\caption{Comparison of the price trajectories of financial securities and the motion of schooling fish. (A) Price trajectories of 40 large-cap US stocks from 1pm to 4pm on February 25, 2010 (from the NASDAQ dataset).  The mean trajectory is shown in black.  Price trajectories are the change in logarithmic price from 1pm to the current time measured every 30 seconds and are smoothed by taking a moving average with window size 20.  The plot is enlarged from 3:02pm to 3:25pm with the following stocks in color: AAPL (red), AMGN (cyan), CMCSA (magenta), HON (yellow), CELG (blue), MOS (dark green), BHI (light green), SWN (pink).  (B) Tracked motion of eight schooling mosquitofish.  The image consists of two superimposed frames at 15s and 17s of the movie provided as supplemental material in\cite{HerbertRead11}.
}
\end{figure*}

With this simple model, synchronizing the prices of securities (allowing the price of security $j$ to affect the price of security $i$) lowers transaction costs and increases the accuracy of prices.  See the supplemental material for a full proof\cite{Materials}.  Here, I show the result with an example. Assume that the market contains only two securities, $n=2$, and the initial price of each security is $m_{1}=m_{2}=50$ which can increase or decrease by $\delta_1=\delta_2=1$.  For both securities $\phi_1=\phi_2=0.75$, and their price changes are correlated with $\rho_{1,2}=0.8$.  The diagram in Fig.~3 analyzes the expected transaction cost, $c(B_1) = E[p_1-m_1|B_1]$, and average pricing error, $e(B_1)=E[|x_1-p_1||B_1]$, for a buy order in security 1, when prices are and are not synchronized (see the supplemental material for details\cite{Materials}).  As seen in the diagram, by allowing order flow in security 2 to affect the price of security 1, a buy order for security 1 costs less and is priced more accurately.

When prices are synchronized, transaction costs are lower in the model.  For this reason, average investors -- individuals who do not correlate their orders with final price changes -- do not lose as much money.  Who, then, is compensating average investors?  When prices are synchronized, information diffuses rapidly from security to security and informed investors who trade different but related securities are forced to compete with one another.  They make less profit as a result.  In the above example, their average profit is reduced from 0.5 to 0.46875 per transaction\cite{Materials}.

Price synchronization is normally beneficial in markets, but it also can have harmful effects.  If shared misconceptions exist in the population of investors within the model -- causing for example, a large number of sell orders to be submitted even though the final prices of most securities are higher -- then synchronization makes transaction prices less accurate.  In addition, when prices are tightly connected to one another, errors quickly spread through the financial system.  To mitigate this risk, HFT firms can program their systems to exit the market when errors are detected.  But determining the difference between an extreme event and an error is precisely the type of problem that machines find difficult\cite{Mueller06}.  Machines that continually stay in the market risk propagating errors when they arise.  Machines that leave the market at the first sign of an abnormality will often disappear at the precise time they are most needed.  The end result is a financial system that becomes unstable during times of stress and behaves very much like US markets did during the Flash Crash\cite{Kirilenko11}.

\section{HFT Activity}

Although HFT firms synchronize prices, this does not imply it is their main activity.  Just how important is synchronization to HFT?  In Fig.~4A, I plot the relationship between the level of HFT activity within a stock and the correlation strength of that stock to other stocks.  Correlations are between the 30 second returns of each stock and the equal-weighted average 30 second returns of all 120 stocks.  HFT activity is measured as the fraction of overall share volume attributable to HFT for each stock.  The correlation between these variables is 0.80, and the $R^2$ from a linear fit is 0.64.  HFT activity varies significantly from security to security, and synchronization explains the majority of this variance.

To determine if HFT is enforcing plausible economic relationships between securities, I calculate the minimum spanning tree of the correlation network\cite{Mantegna99} for the 40 large-cap stocks (Fig.~4B).  The ticker for each stock is shown on the corresponding node, and nodes are color-coded according to their GICS (Global Industry Classification Standard) sector.  The correlation structure of these stocks at 30 second intervals -- largely set by HFT -- corresponds well with the economic relationships of the companies.

Most HFT firms are run by scientists and engineers, and it is unlikely that they pay close attention to economic fundamentals and create a map of market structure that updates as fundamentals change.  Instead, it is more likely that HFT firms are dependent on feedback mechanisms that punish them when the structure they enforce is incorrect.  If and how this feedback mechanism works is an important area of future research.

\section{Analogy to Animal Groups}

It is interesting to compare the above results to recent findings in ecology.  In animal groups, synchronized behavior facilitates information transfer between individuals, which increases the accuracy of decisions and allows fewer resources to be allocated to information gathering\cite{Couzin05, Conradt11}.  A simple example is a school of fish.  By synchronizing their behavior, fish can scan their environment using ``many eyes'', which allows them to quickly evade threats or move towards potential food sources.  

Financial markets are similar.  In markets, the state of the economy is monitored by a large number of investors who quickly broadcast any changes to each other and the rest of society via price movements\cite{Hayek45, Fama70}.  By synchronizing prices, HFT allows the ``many eyes'' of different investors to function as one coherent group, which results in price trajectories that look like the motions of schooling fish (Fig.~5).

Just as in animal groups, synchrony in financial markets leads to an increase in efficiency and a reduction in the resources spent on informed individuals.  However, also as in animal groups, synchronization can have harmful effects; shared misconceptions among individuals in a group are amplified when behaviors are synchronized\cite{Conradt11, Sumpter09}.

\section{Conclusions}

The evidence above suggests that HFT plays an important role in financial markets.  By synchronizing prices, HFT facilitates information transfer between investors, which increases the accuracy of prices and redistributes profits from informed individuals to average investors by reducing transaction costs.

Synchronization, however, is not a panacea for markets.  When prices are tightly connected to one another, errors can quickly propagate throughout the financial system if safeguards are not in place.  In addition, if shared misconceptions exist among investors, they are amplified so that prices are less accurate overall.  Finally, synchronization can create spurious structure in markets if information about the changing relationships of securities does not make its way to the high-frequency domain. 

In sum, these results help answer several puzzling questions about HFT that previously seemed difficult to explain: (1) why HFT is so prevalent, (2) why HFT increases market efficiency under normal conditions but leads to instability during times of stress, (3) why HFT concentrates in certain securities and not in others, and finally, (4) how HFT can lower transaction costs yet still make profits.

\begin{acknowledgments}
I thank J. D. Farmer, K. Glover, D. Michayluk, and F. Reed-Tsochas for helpful comments and suggestions. This work was supported by the European Commission FP7 FET-Open Project FOC-II (no. 255987).
\end{acknowledgments}

\end{document}